\documentclass[pre,twocolumn,superscriptaddress,showpacs]{revtex4-1}

\usepackage[dvipdfmx]{graphicx}
\usepackage{amsmath,amssymb}
\usepackage{dcolumn}
\usepackage{bm}
\usepackage{color}
\usepackage{hyperref}
\hypersetup{colorlinks=true,breaklinks,linkcolor=blue,urlcolor=blue,citecolor=blue}

\def\mathi{\mathrm i}

\begin{document}

\preprint{}

\title{Sparse modeling approach to analytical continuation\\ of imaginary-time quantum Monte Carlo data}

\author{Junya Otsuki}
\affiliation{Department of Physics, Tohoku University, Sendai 980-8578, Japan}
\author{Masayuki Ohzeki}
\affiliation{Graduate School of Information Sciences, Tohoku University, Sendai 980-8579, Japan}
\author{Hiroshi Shinaoka}
\affiliation{Department of Physics, Saitama University, 338-8570, Japan}
\author{Kazuyoshi Yoshimi}
\affiliation{Institute for Solid State Physics, University of Tokyo, Chiba 277-8581, Japan}

\date{\today}

\begin{abstract}
A new approach of solving the ill-conditioned inverse problem for analytical continuation is proposed.
The root of the problem lies in the fact that even tiny noise of imaginary-time input data has a serious impact on the inferred real-frequency spectra.
By means of a modern regularization technique,
we eliminate redundant degrees of freedom that essentially carry the noise,
leaving only relevant information unaffected by the noise.
The resultant spectrum is represented with minimal bases and thus a stable analytical continuation is achieved.
This framework further provides a tool for analyzing to what extent the Monte Carlo data need to be accurate to resolve details of an expected spectral function.
\end{abstract}



\maketitle

\section{Introduction}
Numerical and analytical calculations in quantum many-body systems are in most cases performed with the imaginary-time framework.
Diagrammatic perturbation theory~\cite{AGD} and variants of quantum Monte Carlo (QMC) simulations~\cite{Gull11,Gubernatis-book} take full advantage of imaginary-time descriptions of statistical averages.
One needs, however, to perform analytical continuation to transform calculated imaginary-time quantity $G(\tau)$ to real-frequency spectra $\rho(\omega)$, which can be compared directly to experimental results.
This becomes problematic particularly when handling QMC data, because analytical continuation is
extremely sensitive to noise.

Due to the sensitivity to noise, 
the standard Pad\'e approximation~\cite{Vidberg77} often yields unphysical spectra that even break preconditions such as the sum rule and causality.
In order to stably obtain a physically reasonable spectrum, various numerical algorithms have been developed such as the maximum entropy method (MaxEnt)~\cite{MaxEnt,Gunnarsson10b,Bergeron16} and stochastic method~\cite{Sandvik98,Mishchenko00,Fuchs10,Beach-arXiv,Sandvik16}.
With the recent progress in computational and information theories, 
there are yet growing attempts to settle this long-standing issue~\cite{Beach00,Oestlin12,Dirks13,Bao16,Schoett16,Levy-arXiv,Goulko17,Bertaina16,Arsenault-arXiv}.
Despite extensive efforts from many different angles, a fundamental question still remains: to what extent imaginary-time data with statistical errors have relevant information in the first place, and in other words, how much we can, in principle, reconstruct fine structure of real-frequency spectra.

In this paper, we address this fundamental issue by presenting
a new approach based on the concept of sparse modeling (SpM), which has been developed in the context of data-driven science.
Technically, the SpM provides ways to extract relevant variables for representing high-dimensional data, eliminating redundant variables that potentially cause overfitting.
Our idea is that, by {\em enforcing sparseness on imaginary-time data represented in a properly constructed basis set}, we are able to extract relevant information to perform stable analytical continuations against noise.
We shall demonstrate that this idea does work and further reveals
the accuracy of imaginary-time QMC data required for reproducing structure of spectral functions.

\section{Formalism}
\subsection{Descriptions of the problem}
The input of analytical continuation is the imaginary-time Green function $G(\tau)$ or its Fourier transform $G(\mathi\omega_n)$, where $\omega_n$ denotes a Matsubara frequency.
If the analytical expression of $G(\mathi\omega_n)$ is known,
one can readily obtain the spectral function $\rho(\omega)$ using the relation $\rho(\omega)=(1/\pi) {\rm Im}G(\omega+\mathi0)$ by replacing $\mathi \omega_n$ with $\omega+\mathi0$.
For numerical data,
one may use the exact integral equation between $\rho(\omega)$ and $G(\tau)$~\footnote{Note the sign of $G(\tau)$: In our definition, $G(\tau)$ is positive definite in $0\leq\tau\leq\beta$.}
\begin{align}
G(\tau) = \int_{-\infty}^{\infty} d\omega K_{\pm}(\tau, \omega) \rho(\omega),
\label{eq:G_K_rho}
\end{align}
where $0\leq\tau\leq\beta\equiv1/T$ and the kernel $K_{\pm}$ is given by
\begin{align}
\label{eq:kernel}
K_{\pm}(\tau, \omega)=\frac{e^{-\tau\omega}}{1 \pm e^{-\beta\omega}}.
\end{align}
Here, the $+$ ($-$) sign is for fermionic (bosonic) correlation functions.
In this representation, the analytical continuation may be read as an inverse problem in which one infers $\rho(\omega)$ from its integrated values $G(\tau)$ in the presence of noise.
This constitutes an ill-posed problem: the kernel $K$ is exponentially small at large $\omega$,
and so the right-hand side of Eq.~(\ref{eq:G_K_rho}) is insensitive to variations of $\rho(\omega)$.
Therefore, tiny noise in the left-hand side can considerably affect the ``best'' inference of $\rho(\omega)$.
In other words, there are enormous number of plausible solutions that satisfy Eq.~(\ref{eq:G_K_rho}) within a given accuracy.
Our objective is to {\em select} a reasonable solution which is independent of noise.

For convenience sake, we recast Eq.~(\ref{eq:G_K_rho}) into a conventional linear equation with dimensionless quantities as
\begin{align}
\label{eq:y_Kx}
\bm{G} = K \bm{\rho}.
\end{align}
Here, the vector $\bm{G}$ is defined by $G_i \equiv G(\tau_i)$ with $\tau_i$ being $M$-division of $[0:\beta]$. 
In the fermionic cases, the quantities on the right-hand side are defined by $K_{ij} \equiv K_+(\tau_i, \omega_j)$ and $\rho_j \equiv \rho(\omega_j)\Delta\omega$~\footnote{In bosonic cases, $K$ and $\bm{\rho}$ are defined by $K_{ij} \equiv \omega K_-(\tau_i, \omega_j)$ and $\rho_j \equiv (\rho(\omega_j)/\omega)\Delta\omega$. Then, all descriptions below are applicable.}, which are obtained after 
replacing the integral over $\omega$ with $N$-point finite differences in the range $[-\omega_{\rm max}:\omega_{\rm max}]$.
One may use a non-linear mesh for better efficiency, but the formulation below does not change.
When the input $\bm{G}$ has noise, 
deviation from Eq.~(\ref{eq:y_Kx}) needs to be taken into account.
For this reason, 
we consider the square error
\begin{align}
\chi^2(\bm{\rho}) = \frac12 \| \bm{G} - K \bm{\rho}\|_2^2,
\label{eq:MSE}
\end{align}
and find $\bm{\rho}$ such that $\chi^2(\bm{\rho}) <\eta$ with $\eta$ being a small constant depending on the magnitude of noise.
Here, $\| \cdot \|_2$ stands for the $L_2$ norm defined by $\| \bm{\rho} \|_2 \equiv (\sum_j \rho_j^2 )^{1/2}$.

The solution must hold two conditions: non-negativity $\rho(\omega) \geq 0$ and the sum rule $\int_{-\infty}^{\infty} \rho(\omega) d\omega =1$.
They are expressed in terms of the vector $\bm{\rho}$ as
\begin{align}
\label{eq:constraint_x}
\rho_j \geq 0,\quad \sum_j \rho_j = 1.
\end{align}
These constraints are applied to diagonal components of Green functions.
For off-diagonal components, the non-negativity is not applied, while the sum rule always exists~\footnote{We can determine the spectral sum $c$ by $c=G(\tau=0)+G(\beta)$ or by the high-frequency tail $-c/(i\omega_n)$.}.
Our algorithm presented below works both with and without those constraints.
This is a technical advantage over MaxEnt, in which the entropy term requires the positiveness, $\rho(\omega)>0$~\footnote{For non-negative spectra in MaxEnt, see Ref.~\cite{Reymbaut15}.}.

\subsection{Efficient basis set}
We discuss what basis best describes spectral functions $\rho(\omega)$.
Here, the ``best'' means ability of reproducing the correct $\rho(\omega)$ (i) with a small number of bases (ii) for wide models independent of details of interactions/parameters.
For this purpose, we focus on the fact that the matrix $K$ is ill-conditioned.
To see this, 
we use the singular value decomposition (SVD) of the matrix $K$:
\begin{align}
\label{eq:SVD}
K = U S V^{\rm t},
\end{align}
where $S$ is an $M\times N$ diagonal matrix, and $U$ and $V$ are orthogonal matrices of size $M\times M$ and $N\times N$, respectively.
It should be noted that the singular values $s_l$ ($l=0,1,2,\cdots$) decay exponentially or even faster [see Fig.~\ref{fig:svd_basis}(a1)].
This makes a numerical optimization of $\chi^2(\bm{\rho})$ unstable.
A standard recipe for avoiding this difficulty is to drop vectors corresponding to small singular values below a certain threshold~\cite{NumericalRecipes}\footnote{This trick is known as Bryan algorithm in the context of MaxEnt~\cite{MaxEnt}.}.
Although this yields some definite solution, the result depends totally on the threshold.
We make another use of SVD of the ill-conditioned matrix in modern perspective of SpM.

\subsection{$L_1$ regularization}
We reconsider the expression of $\chi^2(\bm{\rho})$ in Eq.~(\ref{eq:MSE}).
Introducing new vectors
\begin{align}
\label{eq:SV_basis}
\bm{\rho}'\equiv V^{\rm t}\bm{\rho},
\quad
\bm{G}'\equiv U^{\rm t}\bm{G},
\end{align}
we obtain
\begin{align}
\chi^2(\bm{\rho}') = \frac{1}{2} \| \bm{G}' - S \bm{\rho}'\|_2^2 = \frac{1}{2} \sum_l (G'_l - s_l \rho'_l)^2.
\label{eq:MSE-2}
\end{align}
It turns out that the contribution of $\rho'_l$ to $\chi^2(\bm{\rho}')$ is weighted by the corresponding singular value $s_l$. 
Since $s_l$ decays exponentially as noted above, most elements of $\bm{\rho}'$ give only negligible contribution to $\chi^2(\bm{\rho}')$.
Hence, such elements are essentially indefinite as far as $\chi^2(\bm{\rho}')$ is concerned, making a naive analytical continuation quite sensitive to noise.

The above consideration
brings us to the idea that, by imposing sparseness on $\bm{\rho}'$, we can find a stable solution
which is robust against noise.
To this end, 
we consider the cost function including an $L_1$ regularization term
\begin{align}
\label{eq:F}
F(\bm{\rho}') \equiv \frac12 \| \bm{G}' - S \bm{\rho}'\|_2^2 + \lambda \| \bm{\rho}' \|_1,
\end{align}
where $\lambda$ is a positive constant
and $\| \cdot \|_1$ denotes the $L_1$ norm defined by
\begin{align}
\label{eq:L1}
\| \bm{\rho}' \|_1 \equiv \sum_l |\rho_l'|.
\end{align}
This form of optimization problems is referred to as LASSO (Least Absolute Shrinkage and Selection Operators)~\cite{Tibshirani96}.

\begin{figure}[tb]
	\begin{center}
	\includegraphics[scale=0.9]{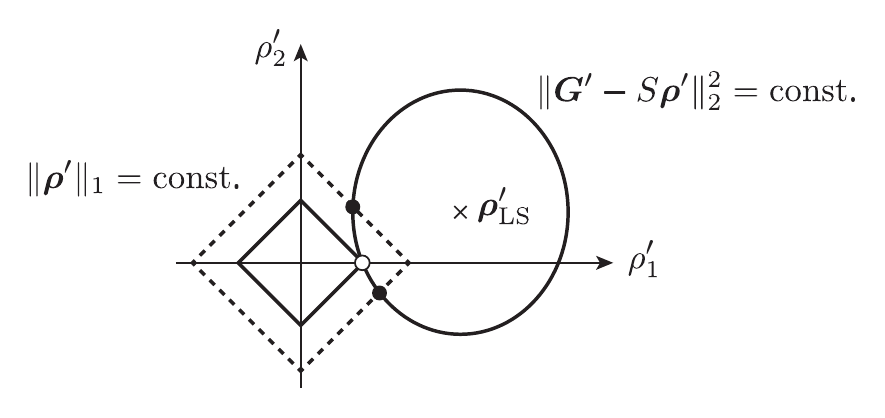}
	\end{center}
	\caption{Explanation for the mechanism that a sparse solution is chosen by the $L_1$ regularization. For details, see the paragraph below Eq.~(\ref{eq:L1}).}
	\label{fig:L1}
\end{figure}

The role of the $L_1$ term may be explained as follows. We consider a two-dimensional vector $\bm{\rho}'=(\rho'_1, \rho'_2)$ as the simplest example (Fig.~\ref{fig:L1}). 
We suppose that the minimum of $\chi^2(\bm{\rho}')$, namely, the solution of the least-square method, is located at $\bm{\rho}'=\bm{\rho}'_{\rm LS}$.
Equal-value contours of $\chi^2(\bm{\rho}')$ are elliptic centered at $\bm{\rho}'_{\rm LS}$, and all points on this line, e.g. open and closed circles in Fig.~\ref{fig:L1}, are equally feasible when a certain extent of errors are allowed.
When the $L_1$ term is included, the open circle becomes the most favorable, since the contours of the $L_1$ norm exhibit a cusp on each axis.
The $L_1$ term thus selects a sparse solution out of an infinite number of plausible solutions of least squares.
Furthermore, since the LASSO is a convex optimization, the global minimum can be obtained regardless of initial conditions~\cite{Boyd-book}. Therefore, the present scheme is computationally inexpensive.

Our task now is to find $\bm{\rho}'$ that minimizes Eq.~(\ref{eq:F}) subject to the constraints in Eq.~(\ref{eq:constraint_x}).
We applied an algorithm named alternating direction method of multipliers (ADMM) developed by Boyd {\it et al.}~\cite{Boyd11}.
See Appendix~\ref{app:ADMM} for closed explanations for this algorithm.
How to choose a reasonable value of $\lambda$ will be discussed later.

\section{Demonstrative results}
\begin{figure}[tb]
	\begin{center}
	\includegraphics[width=\linewidth]{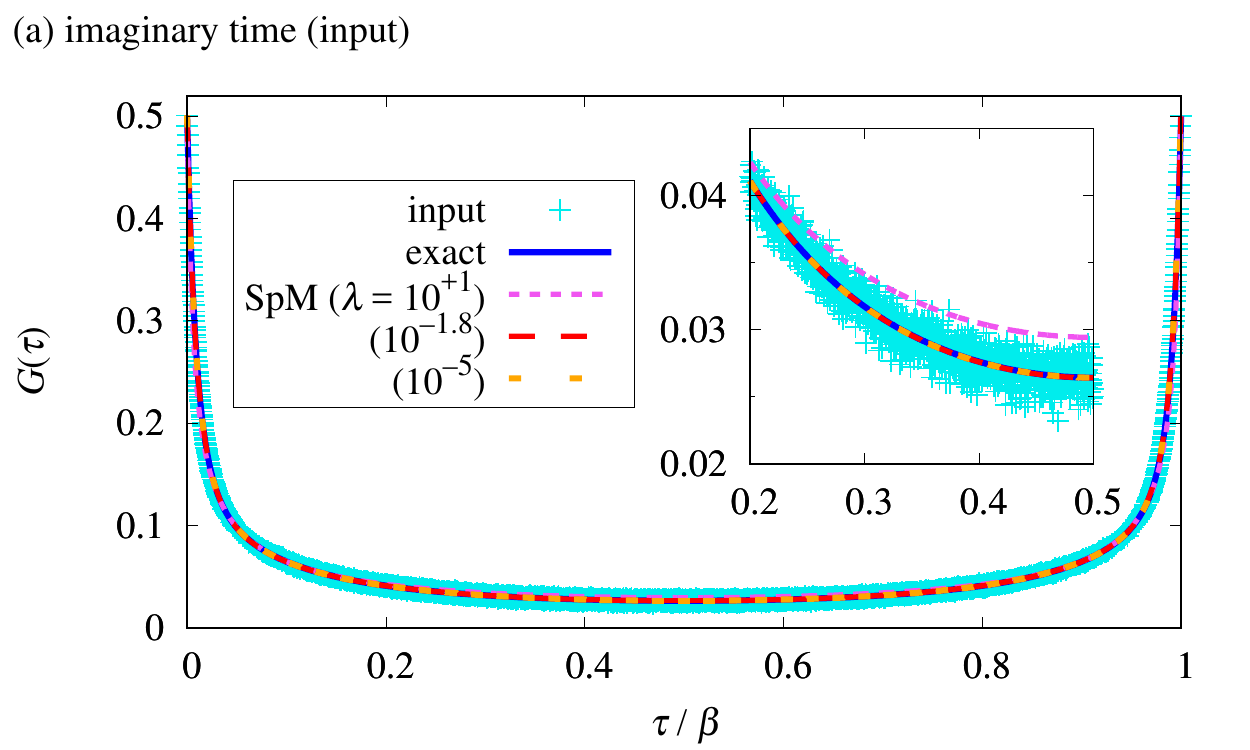}
	\includegraphics[width=\linewidth]{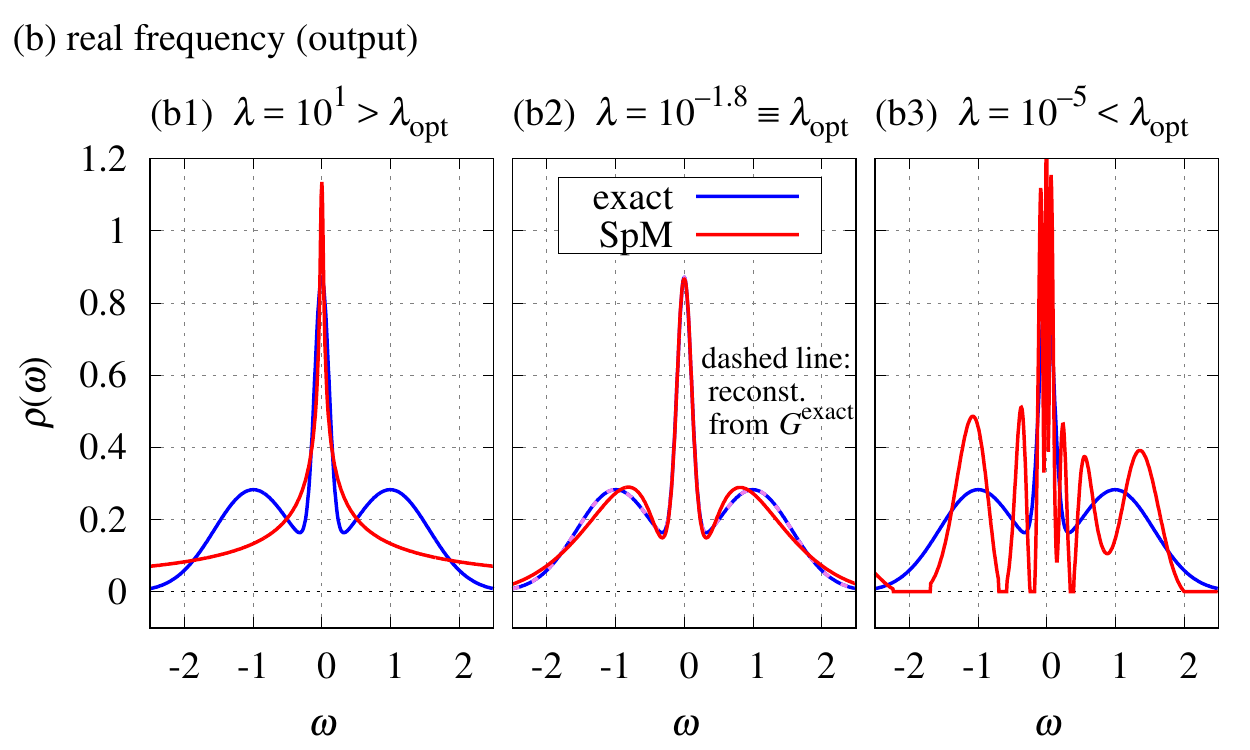}
	\end{center}
	\caption{(a) Three kinds of imaginary-time data: the exact $G^{\rm exact}$ without noise, the input data $G^{\rm input}$ with noise, and the result $G^{\rm SpM}$ recovered after analytical continuations. (b) Two kinds of real-frequency data: the exact spectrum $\rho^{\rm exact}$ and the spectrum $\rho^{\rm SpM}$ reconstructed from $G^{\rm input}$ using our scheme. Here, $\rho^{\rm exact}$ consists of three Gaussians with parameters $({\rm position, width, weight})=(0, 0.15, 0.2), (\pm1, 0.8, 0.4)$. Three panels, (b1)--(b3), are for different values of $\lambda$. The dashed line in (b2) (almost overlapping with $\rho^{\rm exact}$) shows a spectrum reconstructed from the noise-less input, $G^{\rm exact}$, for comparison ($\lambda=10^{-12}$).}
	\label{fig:spectrum}
\end{figure}

We present demonstrative results of our analytical continuation scheme.
Imaginary-time input data were prepared as follows. 
We construct a model spectrum $\rho^{\rm exact}(\omega)$ with three Gaussians that imitate a typical single-particle excitation spectrum in the single-impurity Anderson model,
as shown in Fig.~\ref{fig:spectrum}(b) (see the caption for details).
This spectrum is transformed into $G^{\rm exact}(\tau)$ by performing the integral in Eq.~(\ref{eq:G_K_rho}) with $\beta=100$. 
Supposing QMC calculations, we introduced Gaussian noise $\bm{\eta}$ with the standard deviation $\sigma=10^{-3}$.
Then, we obtained input data, $G^{\rm input}_i=G^{\rm exact}(\tau_i)+\eta_i$ [Fig.~\ref{fig:spectrum}(a)].
The discretization parameters are $M=4001$ for $\tau$ and $N=1001$ for $\omega$, and  cutoff is $\omega_{\rm max}=4$.

Figure~\ref{fig:spectrum}(b) shows the spectrum $\rho^{\rm SpM}(\omega_j)=\rho_j^{\rm SpM}/\Delta\omega$ reconstructed by SpM.
Results for three different values of $\lambda$ are plotted:
an optimal choice $\lambda= 10^{-1.8} \equiv \lambda_{\rm opt}$ in (b2), and larger and smaller values in (b1) and (b3), respectively.
How to estimate $\lambda_{\rm opt}$ will be discussed later.
In the optimal case, a reasonable agreement is seen around $\omega=0$.
The deviation around $\omega=\pm1$ is due to the noise, since the whole spectrum can be reconstructed in the absence of noise [dashed line in Fig.~\ref{fig:spectrum}(b2)]. 
This deviation indicates the limitation in reconstructing the real-frequency spectrum from the noisy imaginary-time data, which will be discussed later.
The spectrum becomes featureless for strong regularization ($\lambda>\lambda_{\rm opt}$), while artificial spikes appear for weak regularization ($\lambda<\lambda_{\rm opt}$).
The latter is typical overfitting behavior.

\begin{figure}[tb]
	\begin{center}
	\includegraphics[width=\linewidth]{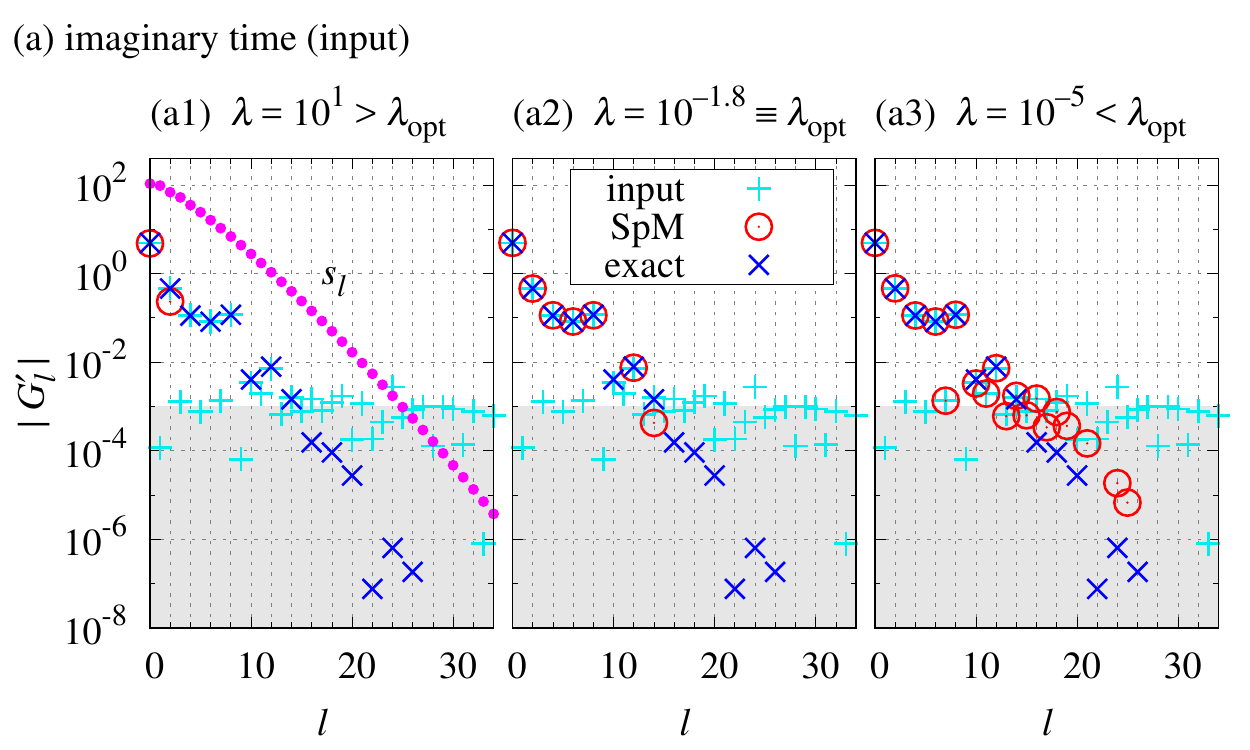}
	\includegraphics[width=\linewidth]{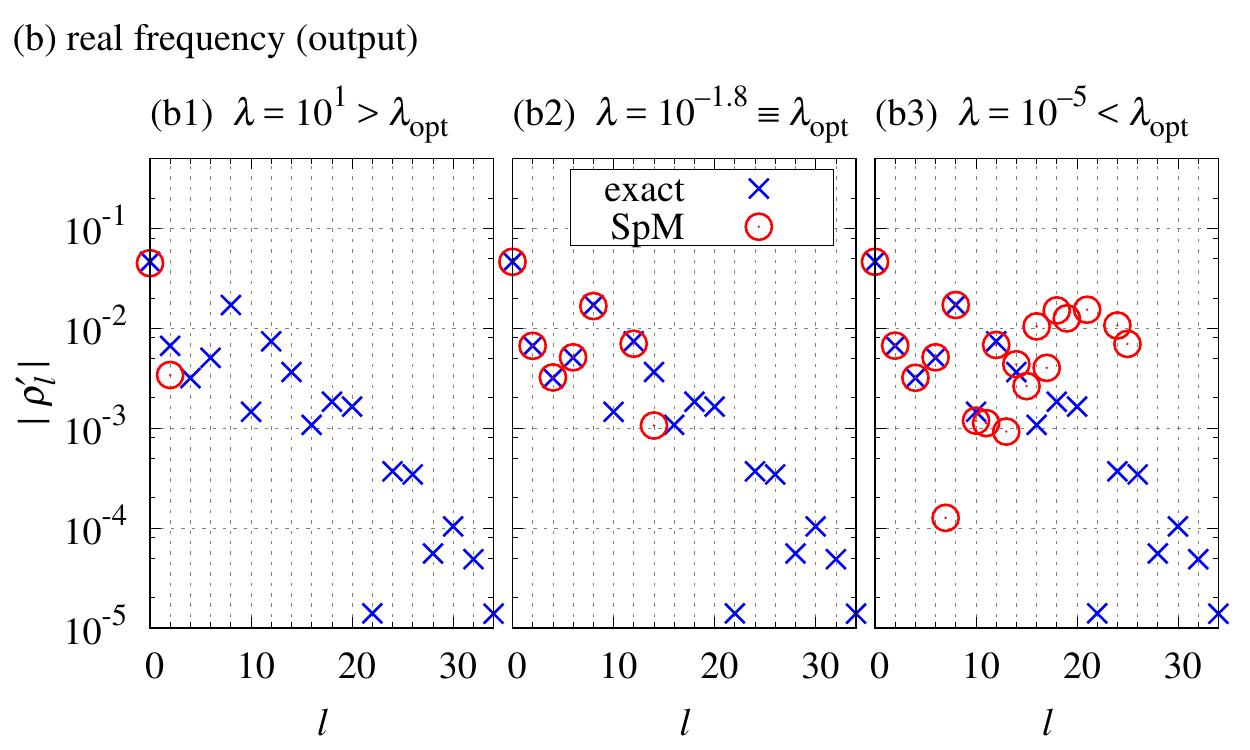}
	\end{center}
	\caption{Imaginary-time ($\bm{G}'$) and real-frequency ($\bm{\rho}'$) data represented in the SV basis corresponding to the ones in $\tau$-$\omega$ basis in Fig.~\ref{fig:spectrum}. The shaded area in (a) indicates regions below $\sigma=10^{-3}$. The closed circles in (a1) show the singular values $s_l$.}
	\label{fig:svd_basis}
\end{figure}

Here, we go back to the imaginary-time data in Fig.~\ref{fig:spectrum}(a) and check agreement between the input $\bm{G}^{\rm input}$ and the SpM result $K\bm{\rho}^{\rm SpM} \equiv \bm{G}^{\rm SpM}$.
In the optimal case, $\bm{G}^{\rm SpM}$ shows a perfect agreement with the exact data without noise rather than $\bm{G}^{\rm input}$, meaning that the noise has been removed.
Note that a equally good agreement is also seen in $\lambda<\lambda_{\rm opt}$, indicating that the two apparently different spectra, Figs.~\ref{fig:spectrum}(b2) and \ref{fig:spectrum}(b3), are equally ``good'' solutions in terms of $\chi^2(\bm{\rho})$.
We will see below that those spectra are clearly distinguished by taking the $L_1$ regularization term into account.

In Fig.~\ref{fig:svd_basis}, the data in Fig.~\ref{fig:spectrum} are represented in the basis defined with SVD in Eq.~(\ref{eq:SV_basis}) (termed as SV basis hereafter).
We first remark that the imaginary-time data in the SV basis, $G'_l$, decay exponentially as shown in Fig.~\ref{fig:svd_basis}(a).
It follows that the input data have only about 6 elements above the magnitude of the noise, $\sigma=10^{-3}$, and other information is lost.
The optimal solution turns out to select those elements properly,
while less (more) elements are selected in the results for $\lambda>\lambda_{\rm opt}$ ($\lambda<\lambda_{\rm opt}$).
Figure~\ref{fig:svd_basis}(b) plots the corresponding real-frequency data, $\rho'_l$.
We find that the spectra are represented with 2 elements for $\lambda>\lambda_{\rm opt}$, 7 elements for $\lambda=\lambda_{\rm opt}$, and many elements including incorrect values for $\lambda<\lambda_{\rm opt}$.

The data in Fig.~\ref{fig:svd_basis}(a) are highly suggestive.
As pointed out above, only a few elements of $\bm{G}'$ possess relevant information unaffected by noise.
This is an intrinsic feature of imaginary-time quantities as explained below.
In the SV basis, Eq.~(\ref{eq:y_Kx}) may be written as $G'_l=s_l \rho'_l$.
Hence, the fast decay of $G'_l$ originates in $s_l$ rather than $\rho'_l$, meaning that it does {\em not} depend on particular models.
It follows that large-$l$ components of $G'_l$ are inevitably buried in noise~\footnote{A similar feature was observed when expanding $G(\tau)$ in terms of the Legendre polynomials~\cite{Boehnke11}.}.
Since large-$l$ bases correspond to highly oscillatory functions,
fine structure of (unknown) exact $\rho(\omega)$ are essentially lost in imaginary-time QMC data.
The present regularization scheme extracts the full information of the input $G(\tau)$, which, however, gives only limited information of the real-frequency counterpart. 

\begin{figure}[tb]
	\begin{center}
	\includegraphics[width=\linewidth]{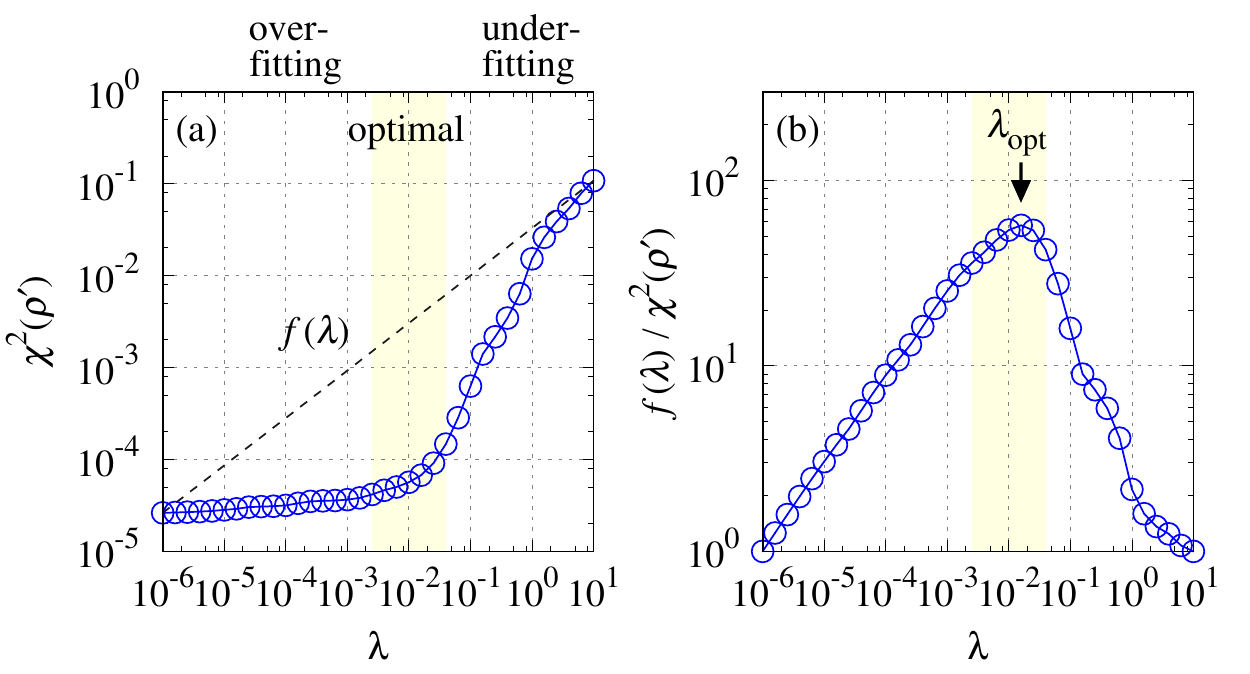}
	\end{center}
	\caption{(a) The square error $\chi(\bm{\rho}')$ as a function of $\lambda$. (b) Reduction of $\chi(\bm{\rho}')$ relative to $f(\lambda)$ shown in (a).}
	\label{fig:lambda_dep}
\end{figure}

Here, we discuss how to find an optimal value of the regularization parameter $\lambda$.
Figure~\ref{fig:lambda_dep}(a) shows $\lambda$ dependence of the square error $\chi^2(\bm{\rho}')$ in Eq.~(\ref{eq:MSE-2}). 
As $\lambda$ decreases from the strong regularization regime, $\chi^2(\bm{\rho}')$ first drops rapidly and then becomes more or less saturated below $\lambda \sim 10^{-2}$, which is an indication of overfitting. 
We did obtained reasonable spectra in a wide region around the kink, namely, $\lambda \simeq 10^{-2.6}$--$10^{-1.4}$ (colored area in Fig.~\ref{fig:lambda_dep}).
Hence, an optimal value may be determined as follows.
We first define a function $f(\lambda)=a\lambda^b$ (a line in log-log scale)
which connects the left and right endpoints of $\chi^2(\bm{\rho}')$ [dashed line in Fig.~\ref{fig:lambda_dep}(a)].
Then, the peak in the ratio $f(\lambda)/\chi^2(\bm{\rho}')$ (difference in log scale) corresponds to the position of the kink in $\chi^2(\bm{\rho}')$ [Fig.~\ref{fig:lambda_dep}(b)]. In this way, we obtained $\lambda_{\rm opt}=10^{-1.8} \approx 1.6\times10^{-2}$.
A similar method was adopted in the literature~\cite{Decelle14, Yamanaka15}.

\section{Required accuracy of QMC data}
\begin{figure}[tb]
	\begin{center}
	\includegraphics[width=\linewidth]{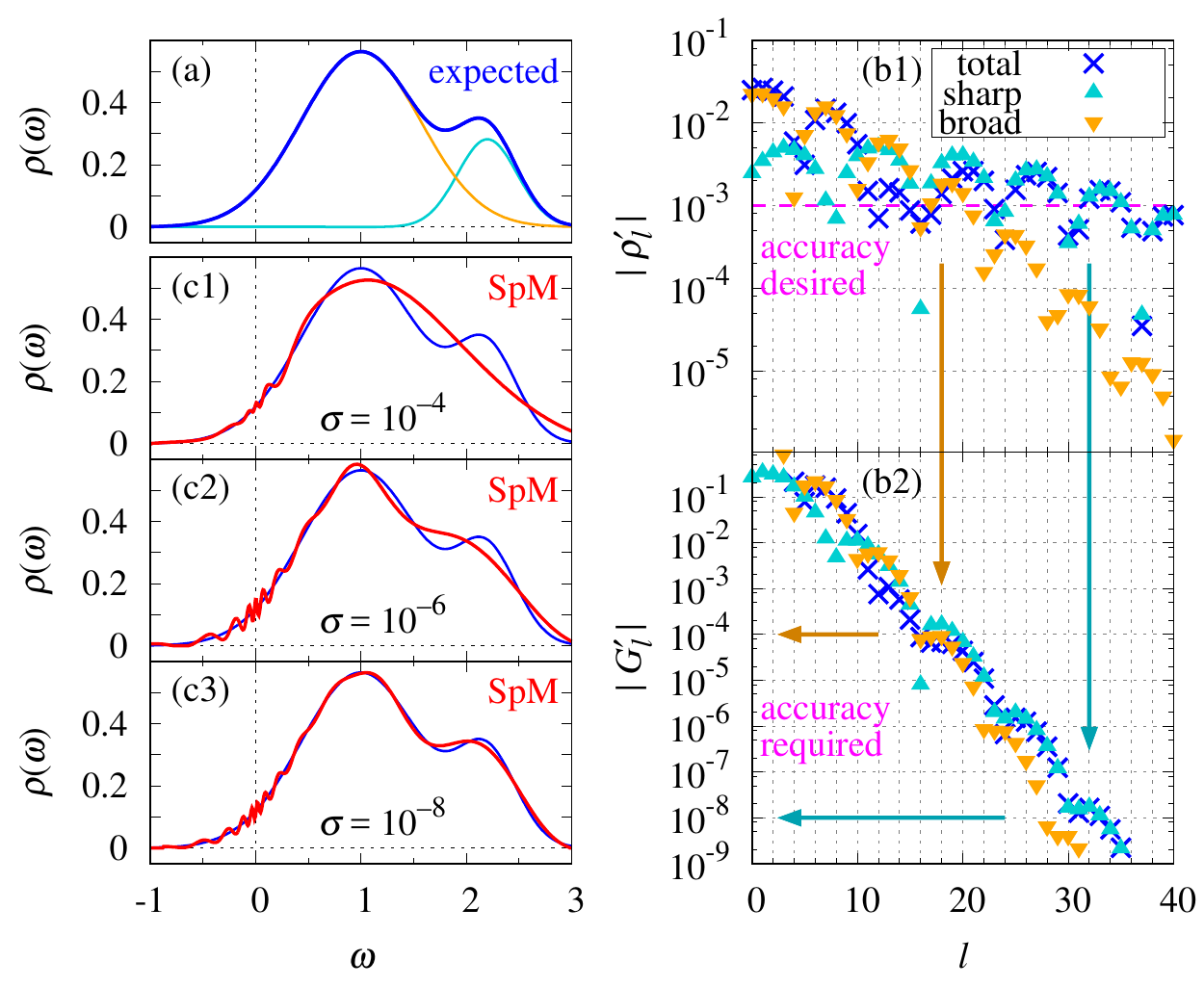}
	\end{center}
	\caption{(a)~A model spectrum $\rho^{\rm expect}(\omega)$ consisting of two Gaussian peaks: $({\rm position, width, weight})=(1, 0.8, 0.8), (2.2, 0.4, 0.2)$. (b)~Corresponding $\rho'_l$ and $G'_l$. For the meaning of the arrows and the dashed line, see the main text. (c)~Spectra computed with the SpM scheme from three sets of $G(\tau)$ with different noise levels, $\sigma=10^{-4}$, $10^{-6}$, and $10^{-8}$.}
	\label{fig:twopeaks}
\end{figure}

We conclude this paper by discussing 
a relation between accuracy of QMC data and capability of reproducing spectrum.
Let us consider a situation where the overall structure of $\rho(\omega)$ is already established in the literature, and its finer structure is controversial.
Practical problems of interest are, for example, a peak-like structure at the edge of a Mott gap~\cite{Nishimoto04}, and spin excitations in the square-lattice Heisenberg model~\cite{DallaPiazza14,Yunoki-unpublished}.
For investigating such issues with QMC, it is highly convenient if one can know accuracy of $G(\tau)$ necessary to achieve reliable $\rho(\omega)$ for a specific problem.
As an example to illustrate our idea,
we consider a two-peak spectrum $\rho^{\rm expect}(\omega)$ shown in Fig.~\ref{fig:twopeaks}(a), and suppose that the existence of the sharper peak located at 
$\omega=2.2$
is a matter of issue.
Our principal interest here is how much accuracy is required for QMC data to verify the existence/absence of the two-peak structure.

We first transform $\rho^{\rm expect}(\omega)$ into $\rho^\prime_l$ in the SV basis.
Figure~\ref{fig:twopeaks}(b1) shows $\rho^\prime_l$, where the contribution of each peak is plotted separately.
The data for the sharper peak decays slower than the broader peak.
In general, a narrower peak at a higher energy yields slower decay.
We consider that each peak can be reconstructed with sufficient accuracy from the components satisfying, e.g. $\rho'_l\gtrsim 10^{-3}$,
which correspond to $l\lesssim 18$ and 32 for the broader and sharper peaks, respectively.
Those boundaries are readily converted into a permissible error in $G(\tau)$: As illustrated in Fig.~\ref{fig:twopeaks}(b1)--(b2), we finally obtain
$10^{-4}\equiv \sigma_{\rm broad}$ and $10^{-8} \equiv \sigma_{\rm sharp}$ as required accuracies to reproduce the broader and shaper peaks, respectively.

Now we test these estimations by SpM calculations.
We prepared three sets of input $G(\tau)$ with different noise levels, $\sigma=10^{-4}$, $10^{-6}$, and $10^{-8}$.
Figures~\ref{fig:twopeaks}(c1)--(c3) show the solution obtained with an optimal value of $\lambda$ for each data.
The result for $\sigma=10^{-4}$ exhibits no separable two-peak structure.
As $\sigma$ decreases, the higher-energy peak grows, and finally at $\sigma=10^{-8}$, the distinct two-peak structure is observed as expected. 
The above estimation of the required accuracy thus turns out to be reasonable.
If QMC data with sufficient accuracy ($\sigma_{\rm sharp}$ in the above example) does not yield the expected structure, then we can conclude its absence.

\section{Summary}
The essence of our algorithm is twofold.
First, the SVD enables an efficient representation of imaginary-time input data and real-frequency spectra.
Second, the $L_1$ regularization selects out bases having relevant information, removing noise automatically.
We can thus perform stable analytical continuations
without any tuning parameters.
Instead of using prior knowledge as in other methods, 
our scheme makes full use of the ill-conditioned nature of the kernel, which has been the source of the problem in analytical continuations,
but it now brings a significant advantage in reducing redundant bases.
With this advantage, it is also possible 
to estimate QMC accuracy that is required to resolve the essential features of a given spectral function. 
It will stimulate future investigations of verifying controversial feature of spectral functions.

Remarkably, our results indicated that imaginary-time data contain much less information than its size in the presence of noise, and hence the data size can be considerably reduced.
This compact representation in the SV basis ($G'_l$) may be used not only for analytical continuations but also for
imaginary-time-based calculations such as diagrammatic expansions and QMC measurements.
This possibility will be pursued in a separate paper~\cite{Shinaoka-arXiv}.

\begin{acknowledgments}
We thank H. Hafermann, P. Werner, and A. Koga for useful comments.
J.O. was supported by JSPS KAKENHI Grant Nos. 26800172, 16H01059 (J-Physics).
M.O. was supported by MEXT KAKENHI Grant No. 25120008, JST CREST and JSPS KAKENHI No. 16H04382.
H.S. was supported by JSPS KAKENHI Grant Nos. 15H05885 (J-Physics), 16K17735.
K.Y. was supported by Building of Consortia for the Development of Human Resources in Science and Technology, MEXT, Japan.
\end{acknowledgments}

\appendix

\section{Alternating direction method of multipliers (ADMM)}
\label{app:ADMM}
In this appendix, we present how to solve the optimization problem including an $L_1$ regularization term and additional constraints.
The problem we need to solve is a minimization of $F(\bm{\rho}')$ in Eq.~(\ref{eq:F}) with respect to $\bm{\rho}'$ subject to the constraints in Eq.~(\ref{eq:constraint_x}).
Following the conventional notation, we change the variables as $\bm{G}\to \bm{y}$ and $\bm{\rho}\to \bm{x}$.
Then, the cost function reads
\begin{align}
\label{S:eq:F}
F(\bm{x}') = \frac12 \| \bm{y}' - S \bm{x}' \|_2^2 + \lambda \| \bm{x}' \|_1,
\end{align}
and the constraints is represented as
\begin{align}
\label{S:eq:constraint_x}
x_j \geq 0,\quad \langle \bm{x} \rangle \equiv \sum_j x_j = 1,
\end{align}
where $\bm{x}=V\bm{x}'$.
The dimension of this optimization problem (size of $\bm{x}'$ and $\bm{y}'$) is given by $L=\min(M,N)$, where $M$ and $N$ are the sizes of $\bm{y}$ and $\bm{x}$, respectively.
Actually, we can reduce $L$ without affecting the result, which will be discussed later.

To solve the optimization problem with multiple constraints, we apply ADMM algorithm developed by Boyd {\it et al.}~\cite{Boyd11}.
Following the ADMM procedure, we introduce auxiliary vectors $\bm{z}$ and $\bm{z}'$, and consider minimization of the function
\begin{align}
\widetilde{F}(\bm{x}', \bm{z}', \bm{z})
&= 
\frac{1}{2\lambda} \| \bm{y}' - S \bm{x}' \|_2^2
-\nu(\langle V\bm{x}' \rangle -1)
\nonumber \\
&+ \| \bm{z}' \|_1
+ \lim_{\gamma\to\infty} \gamma \sum_j \Theta(-z_j),
\label{S:eq:ADMM_cost}
\end{align}
subject to
\begin{align}
\bm{z}'=\bm{x}',
\quad
\bm{z}=V\bm{x}'.
\label{S:eq:ADMM_constraint}
\end{align}
Here, we have used a convention that vectors with prime denote quantities represented in the SV basis (dimension $L$) and those without prime in the original $\tau$--$\omega$ basis (dimension $M$ or $N$).
The sum-rule is imposed by the Lagrange multiplier $\nu$,
and non-negativity is represented by an infinite potential $\gamma$ that acts on negative elements ($\Theta$ is the Heaviside step function).
The auxiliary variable $\bm{z}'$ is in charge of the $L_1$ regularization, and $\bm{z}$ the non-negativity.
The essence of this method is that minimization is performed separately for each vector $\bm{x}'$, $\bm{z}'$ and $\bm{z}$, and their consistency is imposed afterwards gradually.
The ADMM algorithm thus achieves flexibility of handling plural constraints and fast convergence of numerical iterations.

The constraints for the auxiliary variables, Eq.~(\ref{S:eq:ADMM_constraint}), are treated by the augmented Lagrange multiplier method.
We here give a brief description on the treatment of the first constraint, $\bm{z}'=\bm{x}'$.
Two kinds of coefficients play a cooperative role:
(normalized) Lagrange multipliers $\bm{u}'$ which couple with $(\bm{z}'-\bm{x}')$ and a coefficient $\mu'$ of a penalty term $\|\bm{z}'-\bm{x}' \|_2^2$.
The parameter $\mu'$ controls speed of convergence, while $\bm{u}'$ is iteratively updated together with its conjugate variable $\bm{z}'$.
Similarly, we introduce $\mu$ and $\bm{u}$ for the second constraint, $\bm{z}=V\bm{x}'$.

Omitting detailed derivations (we refer readers to Ref.~\cite{Boyd11}), we present below update formulas used in actual computations
(left-going arrows mean substitution):
\begin{subequations}
\label{S:eq:update}
\begin{align}
\bm{x}'
\leftarrow&
\left( \frac{1}{\lambda} S^{\rm t} S
 + (\mu'+\mu) \bm{1} \right)^{-1}
\nonumber\\
&\times
\left( \frac{1}{\lambda} S^{\rm t} \bm{y}'
+ \mu' (\bm{z}' - \bm{u}')
+ \mu V^{\rm t}(\bm{z} - \bm{u})
+ \nu V^{\rm t}\bm{e} \right)
\nonumber \\
&\equiv
\bm{\xi}_1 + \nu\bm{\xi}_2,
\\
\bm{z}' \leftarrow&\ {\cal S}_{1/\mu'} (\bm{x}' + \bm{u}'),
\\
\bm{u}' \leftarrow&\ \bm{u}' + \bm{x}' - \bm{z}',
\\
\bm{z} \leftarrow&\ {\cal P}_+
(V\bm{x}' + \bm{u}),
\\
\bm{u} \leftarrow&\ \bm{u} + V\bm{x}' - \bm{z},
\end{align}
\end{subequations}
where $e_i=1$ and
\begin{align}
\nu= \frac{1-\langle V\bm{\xi}_1 \rangle}{\langle V\bm{\xi}_2 \rangle}.
\end{align}
${\cal P}_+$ is a projection operator onto non-negative quadrant, i.e., ${\cal P}_+ z_j = \max(z_j, 0)$ for each element.
${\cal S}_{\alpha}(\bm{x})$ is the element-wise soft thresholding function, which is defined for each element by
\begin{align}
{\cal S}_{\alpha}(x) = \begin{cases}
x-\alpha & (x>\alpha) \\
0 & (-\alpha \leq x \leq \alpha) \\
x+\alpha & (x<-\alpha)
\end{cases}.
\end{align}
The updates in Eqs.~(\ref{S:eq:update}) are repeated until convergence is reached.
Regarding an initial condition, we may simply set all vectors at zero.


The update formulas in Eqs.~(\ref{S:eq:update}) include matrix-matrix products as well as matrix-vector products.
However, since all matrix-matrix products can be performed before iterations, the computational cost for the updates is quite cheap.

The most costly part is the SVD of the $(M\times N)$-matrix $K$. Once it is done, we convert the input data $\bm{y}$ into the SV basis, $\bm{y}'=U^{\rm t}\bm{y}$, and perform the rest of calculations in this representation. At this stage, we can safely drop bases having small $s_l$ of order of rounding errors, e.g., $s_l < 10^{-10}$. This does not affect the result, since components of those bases are finally becomes zero by the $L_1$ regularization. In the case of calculations in this paper, the number of bases are reduced from $(M,N)=(4001,1001)$ to $L=50$ by this treatment. 
Therefore, considerable speedup of the iteration can be achieved.

The convergence procedure depends on the values of the penalty parameters, $\mu$ and $\mu'$.
We typically set them at 1--100. 
To get fast convergence, we may vary those values during iteration as discussed in Ref.~\cite{Boyd11}.

\bibliography{JO}

\end{document}